\documentclass{jetpl}

\usepackage{dcolumn}
\newcolumntype{X}{D{.}{.}{23}}

\usepackage{color}
\usepackage[dvips]{graphicx}
\usepackage{bm}

\newcommand{\half}{{\textstyle{1\over 2}}}
\newcommand{\threehalf}{{\textstyle{3\over 2}}}
\newcommand{\fivehalf}{{\textstyle{5\over 2}}}
\newcommand{\arrow}{{\Leftrightarrow}}

\twocolumn

\lat

\title{Double--Resonance $\boldsymbol{g}$ Factor Measurements\\
       by Quantum Jump Spectroscopy}

\rtitle{Double--Resonance $\boldsymbol{g}$ Factor Measurements\ldots}

\sodtitle{Double--Resonance $\boldsymbol{g}$ Factor Measurements by
          Quantum Jump Spectroscopy}

\author{Wolfgang Quint$^{+}$,
Behnam Nikoobakht$^{*}$, 
Ulrich D. Jentschura$^*$}

\rauthor{W.\,Quint W., B.\,Nikoobakht B., U.\,D.\,Jentschura U.D.}

\sodauthor{Quint, Nikoobakht, Jentschura}

\address{$^+$Gesellschaft f\"{u}r Schwerionenforschung
(GSI), Planckstra\ss{}e 1, 64291 Darmstadt, Germany\\
$^*$Max--Planck--Institut f\"ur Kernphysik,
Postfach 103980, 69029 Heidelberg, Germany \\
and Institut f\"ur Theoretische Physik,
Universit\"{a}t Heidelberg,
Philosophenweg 16, 69120 Heidelberg, Germany}

\dates{12 November 2007}{$*)$}

\abstract{With the advent of high-precision frequency combs that can bridge
large frequency intervals, new possibilities have opened up for
the laser spectroscopy of atomic transitions. Here, we show that
laser spectroscopic techniques can also be used to determine the
ground-state $g$ factor of a bound electron: Our proposal is based
on a double-resonance experiment, where the spin state of a
ground-state electron is constantly being read out by laser
excitation to the atomic L shell, while the spin flip transitions
are being induced simultaneously by a resonant microwave field,
leading to a detection of the quantum jumps between the
ground-state Zeeman sublevels. The magnetic moments of electrons
in light hydrogen-like ions could thus be measured with advanced
laser technology. Corresponding theoretical predictions are also
presented.}

\PACS{12.20.Ds, 14.60.Cd, 31.30.Jv, 13.40.Em, 06.20.Jr, 31.15.-p}

\begin{document}

\maketitle

Recently, there has been a dramatic progress in the precision
laser spectroscopy of atomic transitions, with uncertainties on
the order of $10^{-14}$ for two-photon transitions in
hydrogen~\cite{FiEtAl2004} and even $10^{-17}$ for ultra-violet
(UV) electric-quadrupole transitions in the mercury
ion~\cite{OsEtAl2006} (statistical effects led to a limitation on
the order of $10^{-16}$ for the evaluation of the latter
measurement). By contrast, microwave measurements of the
bound-electron $g$ factor in hydrogen-like ions
\cite{HaEtAl2000prl,BeEtAl2002prl,VeEtAl2004} have been restricted
to a comparatively low level of accuracy, namely in the range of
$10^{-10}$, where the most accurate values have been obtained for
bound electrons in hydrogen-like carbon $^{12}{\rm C}^{5+}$ and
oxygen $^{16}{\rm O}^{7+}$ (for an introductory reviews on
bound-electron $g$ factors and various related experimental
as well as theoretical
techniques, see~\cite{Qu1995,Be2000,Ka2005}). It is tempting to ask if the
accuracy gap between the two categories of measurements might
leave room for improvement of the $g$ factor determination, as
both measurements investigate the properties of bound electrons.
More specifically, the question arises if the additional
``channels'' provided by laser excitation among the discrete
states of the bound system, and the additional possibilites for
the laser cooling of ions (following the original ideas
formulated in Refs.~\cite{coolingidea,BaEtAl1984}), 
can be used as auxiliary devices to
improve the accuracy of the $g$ factor determination via quantum
jump spectroscopy. We also note that double-resonance techniques
for stored ions have already been shown to open up attractive
experimental possibilities with respect to hyperfine transitions
as well as electronic and nuclear $g$ 
factors~\cite{ItWi1981,WiBoIt1983,MaToWe1998}.

The bound-electron (Land\'{e}) $g_j$ factor for an electron bound
in an ion with a spinless nucleus is the proportionality constant
relating the Zeeman energy $\Delta E$ in the magnetic field $B$
(directed along the $z$ axis) and the Larmor precession frequency
$\omega_{\rm L}$ to the magnetic spin projection $m_j = -\half,
\half$ onto that same $z$ axis. In natural units ($\hbar = c =
\epsilon_0 = 1$), we have
\begin{equation}
\label{def} \Delta E = m_j \, \omega_{\rm L}  = m_j \, g_j \,\mu_{\rm B}
\, B \, ,
\end{equation}
where $\mu_{\rm B} = -e/(2 m_e)$ is the Bohr magneton, expressed in
terms of the electron charge $e$ and the electron mass $m_e$.
Deviations from the Dirac--Breit~\cite{Br1928} prediction $g_j(1S)
= 2 \, ( 1 + 2 \sqrt{1 - (Z\alpha)^2} )/3$ are due to quantum
electrodynamic (QED), nuclear and other effects.

The purpose of this note is to answer the following question:
``Is it possible to apply ultra-high precision atomic laser
spectroscopy to bound-electron $g$ factor measurements?'' Our
answer will be affirmative.

In contrast to the continuous Stern--Gerlach effect~\cite{HeEtAl2000},
and complementing a recent proposal for a high-precision
measurement of the $g$ factor in a highly charged ion~\cite{ShEtAl2006},
the current proposal is based on a Penning trap and will be
studied here in conjunction with the hydrogen-like helium ion
$^4{\rm He}^+$, which seems to be well suited for an experimental
realization in the near future. The magnetic field strength $B$ in
the Penning trap can be calibrated via a measurement of the
cyclotron frequency $\omega_c$ of the trapped ion, and the
Land\'{e} $g$-factor for the bound electron is determined by the
relation
\begin{equation}
\label{rel}
g_j = 2 \, (Z-1) \, \frac{m_e}{m_{\rm ion}} \,
\frac{\omega_{\rm L}}{\omega_c}\,,
\end{equation}
where the electron-ion mass ratio $m_e/m_{\rm ion}$ is an
external input parameter and $Z$ is the nuclear charge
number.

In a Penning trap, a single $^4{\rm He}^+$ ion is confined by a
strong homogeneous magnetic field $B$ in the plane perpendicular
to the magnetic field lines and by a harmonic electrostatic
potential in the direction parallel to the field lines
\cite{De1990}. The three eigenmotions of a stored ion are the
trap-modified cyclotron motion (frequency $\omega_+$), the axial
motion (frequency $\omega_z$), and the magnetron motion (frequency
$\omega_-$). The free-space cyclotron frequency $\omega_c = q
B/m_{\rm ion}$ of an ion with charge $q$ can be determined from
the three eigenfrequencies by~\cite{BrGa1982}
\begin{equation}
\omega_c^2 = \omega_+^2 + \omega_z^2 + \omega_-^2.
\label{invariance}
\end{equation}
Experimentally, the eigenfrequencies of the stored ion can be
measured by non-destructive detection of the image currents which
are induced in the trap electrodes by the ion motion. Measurements
on the level of $\delta\omega_c/\omega_c= 7 \times 10^{-12}$ are
achieved \cite{RaThPr2004,DyEtAl2004,ShReMy2005} by careful
anharmonicity compensation of the electrostatic trapping
potential, optimizing the homogeneity and temporal stability of
the magnetic field close to the Penning trap's center, and cooling
the motional amplitudes of the single trapped ion to low
temperatures. Further optimization of experimental techniques
should make it possible to reach an accuracy of (better than)
$\delta\omega_c/\omega_c= 10^{-12}$. In our proposed $g$ factor
measurement, advantage could be taken of the fact that two
frequencies of the same particle are measured simultaneously
(cyclotron vs.~spin-flip), whereas in a mass measurement, the
cyclotron frequencies of two different particles have to be
determined. 

\begin{figure}[htb]
\begin{center}
\begin{minipage}{8.2cm}
\begin{center}
\includegraphics[width=1.0\linewidth]{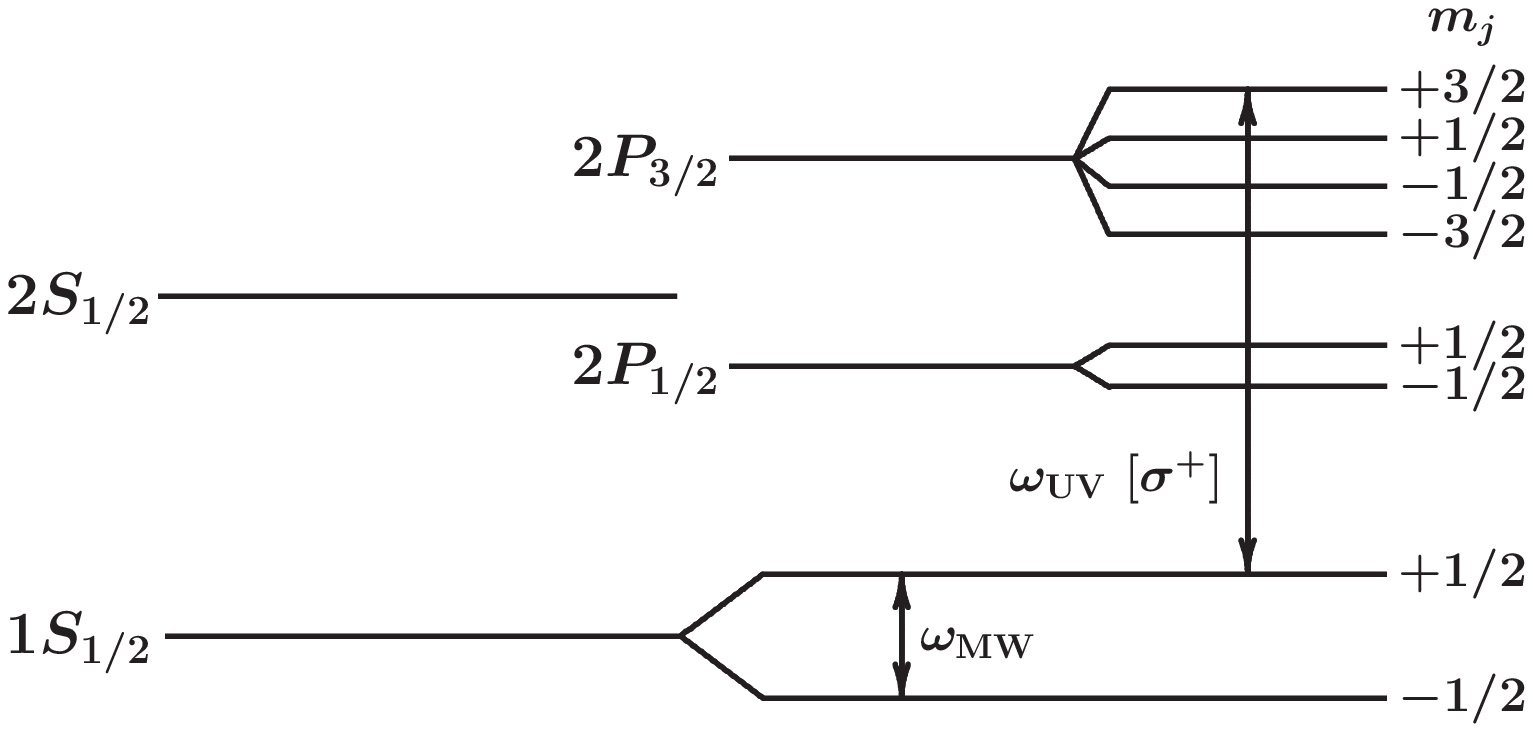}
\caption{\label{fig1} Laser-microwave double-resonance excitation
scheme using circularly polarized UV light for excitation of the
$1S_{1/2}\, \Leftrightarrow 2P_{3/2}$ transition and a microwave
field for driving the spinflip transition $1S_{1/2} \, \left(m_j
\!= \!+\half \right) \Leftrightarrow 1S_{1/2} \, \left(m_j=-\half
\right)$.}
\end{center}
\end{minipage}
\end{center}
\end{figure}

In the following, we concentrate on the $^4 {\rm He}^+$ system,
where the total angular momentum is equal to the total electron
angular momentum $J$. In the presence of the magnetic field $B$ in
the Penning trap, the Zeeman splitting of the electronic ground
state $1S_{1/2}$ of the $^4{\rm He}^+$ ion is given by
Eq.~(\ref{def}). Correspondingly, the excited state $2P_{3/2}$ is
split into four Zeeman sublevels $\Delta E = m_j \, g_j(2P_{3/2})
\, \mu_{\rm B} \, B $, with $m_j= \pm \half, \pm \threehalf$. The
Land\'{e} $g$ factor can be obtained easily 
according to a modified Dirac equation which forms
a basis for bound-state analysis~\cite{EiGrSh2001}
\begin{equation}
g_j(2P_{3/2}) = \frac43 + \frac{\alpha}{3 \pi} - \frac{2}{15} \,
(Z\alpha)^2 \,,
\end{equation}
where we take into account the leading QED and relativistic
contributions and use $Z=2$ for the relativistic term of order
$(Z\alpha)^2$. Suppose now that one single $^4 {\rm He}^+$ ion in
the Penning trap is prepared in the Zeeman sublevel $m_j=+\half$
of the electronic ground state $1S_{1/2}$. Narrow-band ultraviolet
(UV) electromagnetic radiation with $\sigma^+$ polarization and
angular frequency $\omega_{\rm UV} = 2 \pi \times 9.87 \times
10^{15} \, {\rm Hz}$ drives the Lyman-$\alpha$ transition
$1S_{1/2}\, \left(m_j \!= \!+\half \right) \Leftrightarrow
2P_{3/2} \, \left(m_j \!= \!+\threehalf \right)$, see
Fig.~\ref{fig1}. This is a closed cycle because decay by emission
of a fluorescence photon is only possible to the initial state
$1S_{1/2} \, (m_j\!=\!+ \half)$ (if one ignores one-photon
ionization into the continuum). Due to the short lifetime of the
upper state $\tau (2P_{3/2}) \approx 99.7 \, {\rm ps}$, the
fluorescence intensity of $[ 2 \,\, \tau(2p_{3/2}) ]^{-1} \approx
5.01 \times 10^{9} \,\, \mbox{photons}/{\rm s}$ under saturation
conditions makes it possible to detect a single trapped ion with
high sensitivity~\cite{Zscaling}. The Rabi frequency of the UV
transition is given by $\Omega_{\rm Rabi} = 1.308 \times 10^7 \,
{\rm Hz} \, \sqrt{ I_{\rm UV} }$, where $I_{\rm UV}$ is measured
in units of ${\rm W}/{\rm cm^2}$.

Building a continuous-wave (cw) laser that operates at the
Lyman-$\alpha$ transition of $^4{\rm He}^+$, with a wavelength of
$30.37\,{\rm nm}$, is certainly not a trivial task. However, a cw
laser operating at the corresponding Lyman-$\alpha$ transition for
atomic hydrogen, with $\lambda = 121.56\,{\rm nm}$, has already
been demonstrated~\cite{EiWaHa2001}. A possible pathway is
higher-harmonic generation which has recently been described in
Ref.~\cite{GoEtAl2005} and leads to a pulsed UV excitation with a
high repetition rate and a potentially discontinuous probing of
the Zeeman ground-state sublevel. Note that a discontinuous
probing of the ground-state sublevels does not inhibit the quantum
jump detection scheme as outlined below. 
Groundwork for a detailed analysis 
of the dynamics of a pulsed excitation scheme in a very much 
analogous atomic system 
has recently been laid in Ref.~\cite{HaEtAl2006}; in principle,
one only has to ensure that the light intensity of the Lyman-$\alpha$ source 
during a single laser pulse is sufficient to discern the presence
or absence of fluorescence. 

During excitation of the transition $1S_{1/2}\, \left(m_j \!=
\!+\half \right) \Leftrightarrow 2P_{3/2} \, \left(m_j \!=
\!+\threehalf \right)$ and detection of the corresponding
fluorescence photons, a microwave field with frequency
$\omega_{\rm MW}$ in resonance with the spinflip transition
$1S_{1/2} \, \left(m_j \!= \!+\half \right) \Leftrightarrow
1S_{1/2} \, \left(m_j=-\half \right)$ in the electronic ground
state is irradiated on the single trapped $^4 {\rm He}^+$ ion
(Fig.~\ref{fig1}). Successful excitation of the spinflip
transition results in an instantaneous stop of the fluorescence
intensity, because the lower Zeeman level $1S_{1/2}
\left(m_j=-\half\right)$ is not excited by the narrow-band
Lyman-$\alpha$ radiation. A quantum jump is thus directly observed
with essentially 100\% detection efficiency \cite{NaSaDe1986}. A
second spinflip $1S_{1/2} \, \left(m_j = -\half \right) \to
1S_{1/2} \, \left(m_j=+\half \right)$ restores the fluorescence
intensity. A plot of the quantum jump rate versus excitation
microwave frequency at $\omega_{\rm MW} \approx \omega_{\rm L}$
yields the resonance spectrum of the Larmor precession frequency.
The cyclotron frequency $\omega_c$, which also enters
Eq.~(\ref{rel}), is measured simultaneously by non-destructive
electronic detection of the image currents induced in the trap
electrodes \cite{VeEtAl2004}.

\begin{table}[htb]
\begin{center}
\caption{\label{table1} Absolute transition frequencies of $^4{\rm
He}^+$ relevant to the excitation scheme given in Fig.~\ref{fig1}.
The Zeeman splitting is excluded.}
\begin{tabular}{c@{\hspace{0.1cm}}c}
\hline
\hline
\rule[-2mm]{0mm}{6mm}
Nuclear charge radius & $1S \arrow 2P_{3/2}$ frequency\\
\hline
\rule[-2mm]{0mm}{6mm}
$\langle r^2 \rangle^{1/2} = 1.673(1) \, {\rm fm}$ &
$9\,868\,722\,559.240(237)\,{\rm MHz}$ \\
\rule[-2mm]{0mm}{6mm}
$\langle r^2 \rangle^{1/2} = 1.680(5) \, {\rm fm}$ &
$9\,868\,722\,558.650(477)\,{\rm MHz}$ \\
\hline
\hline
\end{tabular}
\end{center}
\end{table}

\begin{figure}[htb]
\begin{center}
\begin{minipage}{8.2cm}
\begin{center}
\includegraphics[width=1.0\linewidth]{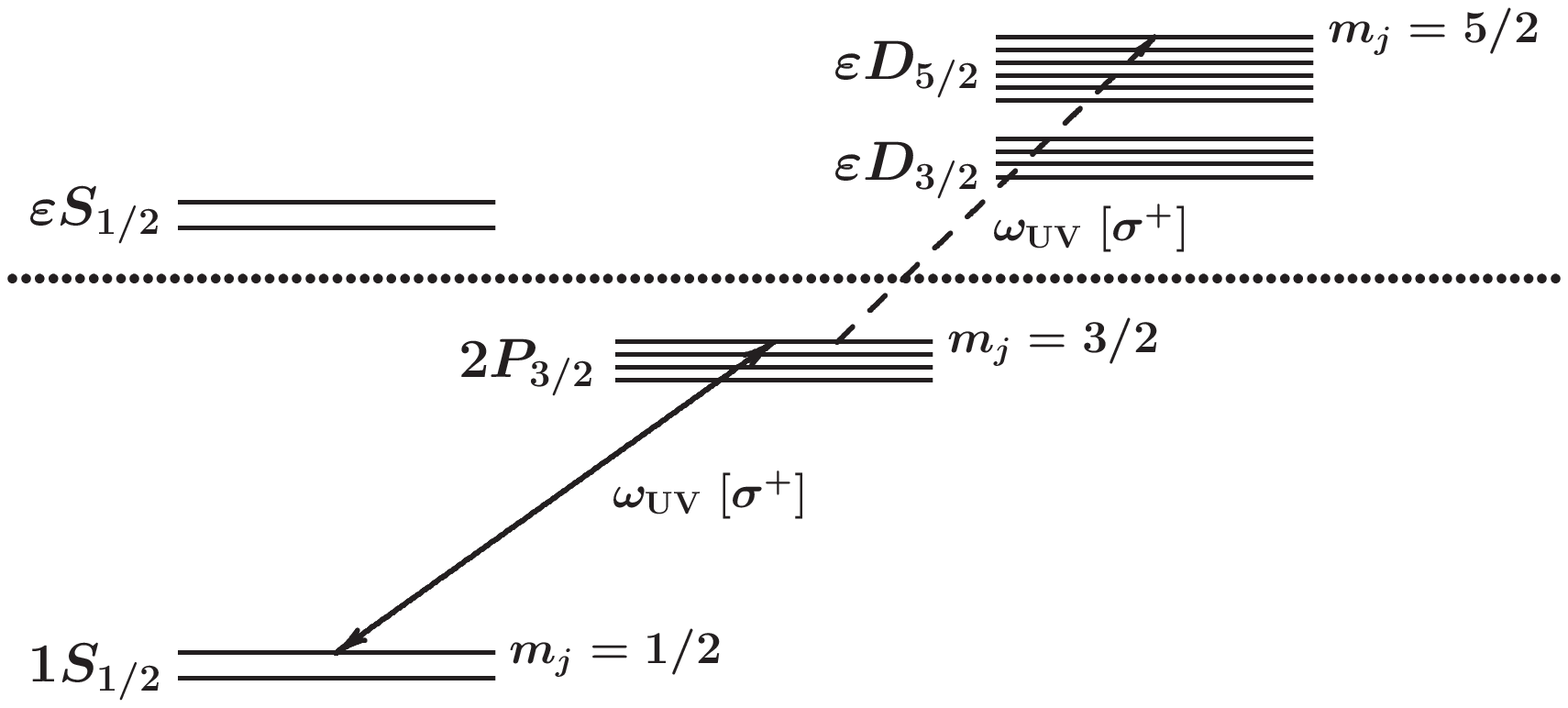}
\caption{\label{fig2} Schematic representation of the excitation
scheme including ionization channels. The dotted line represents
the ionization continuum threshold, and $\varepsilon L_j$ are
electronic continuum states.}
\end{center}
\end{minipage}
\end{center}
\end{figure}

\begin{table*}[htb]
\begin{center}
\begin{minipage}{17cm}
\begin{center}
\caption{\label{table2} Individual contributions to the $1S$
bound-electron $g$ factor for $^4{\rm He}^+$. In the labeling of
the corrections, we follow the conventions of
Ref.~\cite{PaCzJeYe2005}. The abbreviations used are as follows:
``h.o.'' stands for a higher order contribution, ``SE'' for a self
energy correction, ``VP-EL'' for the electric-loop
vacuum-polarization correction, and ``VP-ML'' for the
magnetic-loop vacuum-polarization correction. The value of
$\alpha^{-1} = 137.035\,999\,070(98)$ is the currently most
accurate value from Ref.~\cite{g}, whereas
the value of $\alpha^{-1} = 137.035\,999\,11(46)$ is the 2002
CODATA recommended value~\cite{MoTa2005}.}
\begin{tabular}{c@{\hspace{0.1cm}}cXX}
\hline
\hline
\rule[-2mm]{0mm}{6mm}
$g$ factor contribution ($^4{\rm He}+$) &  &
\multicolumn{1}{c}{$\alpha^{-1} = 137.035\,999\,070(98)$} &
\multicolumn{1}{c}{$\alpha^{-1} = 137.035\,999\,11(46)$} \\
\hline
Dirac eigenvalue & & 1.999~857~988~825~2(2) &
1.999~857~988~825~3(9) \\
Finite nuclear size & & 0.000~000~000~002~3 & 0.000~000~000~002~3 \\
One-loop QED & $(Z\alpha)^{0}$ & 0.002~322~819~466~0(17) &
0.002~322~819~465~4(76) \\
& $(Z\alpha)^{2}$ & 0.000~000~082~462~2 & 0.000~000~082~462~2 \\
& $(Z\alpha)^{4}$ & 0.000~000~001~976~7 & 0.000~000~001~976~7 \\
& h.o.,SE & 0.000~000~000~035~1(2) & 0.000~000~000~035~1(2) \\
& h.o.,VP--EL & 0.000~000~000~002~0 & 0.000~000~000~002~0 \\
& h.o.,VP--ML & 0.000~000~000~000~2 & 0.000~000~000~000~2 \\
$\geq$ two-loop QED & $(Z\alpha)^{0}$ & -0.000~003~515~096~9(3) &
  -0.000~003~515~096~9(3)\\
& $(Z\alpha)^{2}$ & -0.000~000~000~124~8 & -0.000~000~000~124~8 \\
& $(Z\alpha)^{4}$ & 0.000~000~000~002~4(1) & 0.000~000~000~002~4(1)\\
Recoil & $m/M$ & 0.000~000~029~198~5 & 0.000~000~029~198~5 \\
Radiative recoil & $(m/M)^2$ & -0.000~000~000~025~3 & -0.000~000~000~025~3 \\
Hadronic/weak interaction & & 0.000~000~000~003~4 & 0.000~000~000~003~4 \\
Total & & 2.002~177~406~727~1(17) & 2.002~177~406~726~5(77)  \\
\hline
\hline
\end{tabular}
\end{center}
\end{minipage}
\end{center}
\end{table*}

While the UV laser light drives the transition $1S_{1/2}\,
\left(m_j \!= \!+\half \right) \Leftrightarrow 2P_{3/2} \,
\left(m_j \!= \!+\threehalf \right)$, the absorption of an
additional photon can take place, resulting in ionization through
the channel $2P_{3/2} \, \left(m_j \! = \!+\threehalf \right)
\Rightarrow \varepsilon D_{5/2} \, \left(m_j \!= \!+\fivehalf
\right)$, where $\varepsilon D$ are electronic continuum states
(see Fig.~\ref{fig2}).
For this process, we obtain an ionization cross section of $1.631
\, \times 10^{-23} \, {\rm m}^2$. This leads to an ionization
rate of $\gamma_{\rm i} = 2.495 \times 10^{-2} \, {\rm s}^{-1}
\times I_{\rm UV}$, where $I_{\rm UV}$ is the laser intensity
measured in units of ${\rm W}/{\rm cm^2}$, corresponding to a
depletion of the $2P_{3/2} \, \left(m_j=+\threehalf \right)$ state
with a time-dependent exponential $\exp(- \gamma_{\rm i}\, t)$. 
Here, we use the notational conventions of Ref.~\cite{HaEtAl2006}.
In principle, since the ionization rate is proportional to the laser
intensity, whereas the Rabi frequency is only proportional to its
square root, it might be preferable to work at reduced laser
intensities in order to increase the lifetime of the hydrogen-like
charge state of the ion. However, at an incident typical laser
intensity of $100 \, {\rm W}/{\rm cm}^2$, the lifetime of $^4 {\rm
He}^+$ is $0.401\, {\rm s}$ against ionization, and this has to be
compared to a Rabi frequency of $1.308 \times 10^8 \, {\rm Hz}$.
The $^4 {\rm He}^+$ ion has about $10^8$ Rabi cycles before it is
ionized, and so the ionization channel does not limit the
feasibility of the measurement at all.

Finally, we take notice of the ac Stark shift of the $1S$--$2P$
transition due to non-resonant levels, which is $0.0968 \, {\rm
Hz} \times I_{\rm UV}$, with $I_{\rm UV}$ given in ${\rm W}/{\rm
cm}^2$. The ac Stark shift of the UV transition affects the two
ground-state Zeeman levels slightly differently, but the relative
shift of the spinflip transition frequency between them is a
fourth-order effect and is suppressed with respect to the ac Stark
shift by a factor of $\omega_{\rm L}/\omega_{\rm UV} < 10^{-4}$
and thus negligible on the level $10^{-12}$ in units of the
microwave frequency, at a typical laser intensity of $100 \, {\rm
W}/{\rm cm}^2$. Also, experimental procedures for previous $g$
factor measurements~\cite{BeEtAl2002prl,VeEtAl2004} have included
an extrapolation to zero intensity of the microwave fields, and
the same can be done with the driving UV laser field in the
proposed measurement scheme. Alternatively, one can perform the
excitation of the Lyman-$\alpha$ and the spinflip transitions in a
time sequence, thus eliminating any systematic uncertainties of
the $g$ factor determination related to the intensity of the UV
laser light.

In order to lay a theoretical ground for the evaluation of the
$^4{\rm He}^+$ measurement, we present
theoretical predictions for the transition frequencies
and for the $g$ factor.
According to the recent compilations~\cite{vWHoDr2000,JeDr2004,JeHa2007},
the ground state $^4{\rm He}^+$
Lamb shift values are ${\mathcal L}(1S) = 107692.522(228)\,{\rm MHz}$ for
the ``old'' value~\cite{BoRi1978} of the nuclear charge radius
$\langle r^2 \rangle^{1/2} = 1.673(1) \, {\rm fm}$
and ${\mathcal L}(1S) = 107693.112(472)\,{\rm MHz}$ for
the ``new'' value of the charge radius~\cite{SiPriv2006improved}, which is
$\langle r^2 \rangle^{1/2} = 1.680(5) \, {\rm fm}$.
(The uncertainty estimate for the ``old'' value has given
rise to discussions, see Ref.~\cite{vWHoDr2000}.)
For the $2P_{3/2}$ states, the Lamb shift is independent of the current
uncertainty in the nuclear radius on the level of one kHz and
reads ${\mathcal L}(2P_{3/2}) = 201.168\,{\rm MHz}$
(see Tables 3 and 4 of Ref.~\cite{JeHa2007}).
Using a proper definition of the
Lamb shift as given, e.g., in Eq.~(10) of Ref.~\cite{JeHa2007},
we then obtain the transition frequencies
as given in Table~1.

The ground-state $g_j$ factor
can be described naturally in an intertwined expansion in
the QED loop expansion parameter $\alpha$ and the
electron-nucleus interaction strength $Z\alpha$~\cite{PaCzJeYe2005}.
We follow the conventions of Ref.~\cite{PaCzJeYe2005} and
take into account all corrections that are relevant at the $10^{-12}$
level of accuracy (see Table~2).
The entry for the ``$(Z\alpha)^0$ one-loop QED''
is just the Schwinger term $\alpha/(2 \pi)$ and it carries the largest
theoretical uncertainty, because of the uncertainty in the fine-structure
constant $\alpha$ itself~\cite{observation,JeCzPaYe2006}.

In this note, we attempt to formulate a proposal by which 
ultra-accurate $g$ factor measurements 
in hydrogen-like systems with low nuclear charge number might 
be accessible to laser spectroscopic techniques. Within the 
next decade, it is realistic to assume that 
the necessary requirements for experiments will be 
provided that fully profit from both the electric coupling of
the electron (via optical electric-dipole allowed Lyman-$\alpha$
transitions) and from the magnetic coupling of the electron (via
spin-flip transitions among the Zeeman sublevels of the ground
state, see Fig.~\ref{fig1}). The accuracy of the measurement of
the free-electron $g$ factor has recently been increased to a
level of $7.6 \times
10^{-13}$~\cite{DyScDe1987,g}. Within our
proposed setup, an accuracy on the level of $10^{-12} \ldots
10^{-13}$ seems to be entirely realistic for bound-electron $g$
factors in hydrogen-like ions with a low nuclear charge number. It
might be very beneficial if the extremely impressive,
ultra-precise new measurement of the free-electron $g$
factor~\cite{g} could be supplemented by a
potentially equally accurate measurement of the bound-electron $g$
factor in the near future, as an alternative determination of the
fine-structure constant is urgently needed in conjunction with an
improved determination of the electron mass.

The authors acknowledge helpful discussions with 
V. A. Yerokhin and K. Pachucki, and support from DFG
(Heisenberg program) as well as from GSI (contract HD--JENT).


\begin{thebibliography}{10}
\bibitem{Zscaling}
The discussed excitation scheme is in principle applicable to any low-$Z$
  hydrogen-like ion with a spinless nucleus (for higher $Z$, it appears that
  the laser required to drive the $1S$--$2P_{3/2}$ transition is still far from
  experimental possibilities in the foreseeable future). Therefore, it is
  adequate to indicate the scaling of the relevant physical quantities for the
  scheme with the nuclear charge number $Z$. These are as follows: laser
  frequency: $Z^2$; Rabi frequency: $Z^{-1}$ for constant $I_{\rm las}$;
  lifetime of the upper state: $Z^{-4}$; AC Stark shift: $Z^{-4}$, again for
  constant $I_{\rm las}$; ionization rate coefficient: $Z^{-4}$, and ionization
  cross section: $Z^{-2}$. All values given in the text are for $Z=2$.
\end{thebibliography}
\end{document}